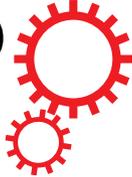

OPEN

# Topological nature and the multiple Dirac cones hidden in Bismuth high-Tc superconductors

Gang Li[1], Binghai Yan[2,3], Ronny Thomale[1] & Werner Hanke[1]

Recent theoretical studies employing density-functional theory have predicted $BaBiO_3$ (when doped with electrons) and $YBiO_3$ to become a topological insulator (TI) with a large topological gap (~0.7 eV). This, together with the natural stability against surface oxidation, makes the Bismuth-Oxide family of special interest for possible applications in quantum information and spintronics. The central question, we study here, is whether the hole-doped Bismuth Oxides, i.e. $Ba_{1-x}K_xBiO_3$ and $BaPb_{1-x}Bi_xO_3$, which are "high-Tc" bulk superconducting near 30 K, additionally display in the further vicinity of their Fermi energy $E_F$ a topological gap with a Dirac-type of topological surface state. Our electronic structure calculations predict the K-doped family to emerge as a TI, with a topological gap above $E_F$. Thus, these compounds can become superconductors with hole-doping and potential TIs with additional electron doping. Furthermore, we predict the Bismuth-Oxide family to contain an additional Dirac cone below $E_F$ for further hole doping, which manifests these systems to be candidates for both electron- and hole-doped topological insulators.

Topological insulators (TIs) are new quantum states of matter that are of fundamental interest for both condensed-matter physics studies and applications in spintronics, quantum information as well as thermoelectrics[3,4]. The unique feature of the TIs is the existence of topologically protected and, therefore, robust conducting channels at the surfaces/edges of materials that are insulating in the interior. Due to the existence of band-inversion in their bulk electronic structure, this new type of insulators is topologically different from the conventional insulators in the sense that TIs cannot be adiabatically transformed into an atomic insulator without going through a phase transition. More precisely, charge conservation and time reversal symmetry around the Fermi level of band inversion establish the protecting symmetry of the non-trivial odd Dirac cone surface state which cannot be gapped out continuously.

Recently, a new direction in the search for topological insulators, with a substantial potential for the above applications, has emerged by identifying $BaBiO_3$ as a TI in the electron-doped region[1,2]. According to density-functional electronic structure calculations[1], this compound possesses the largest topological gap (~0.7 eV) among currently known TI materials and is naturally stable against surface oxidation and degradation, in contrast to other TIs. The large topological gap is induced by the strong spin-orbit coupling (SOC) of Bismuth in cubic $BaBiO_3$, which causes an inversion between the Bi-$s$ and Bi-$p$ band at a time-reversal invariant momenta (TRIM), i.e. the symmetry point R. Inside the corresponding topological gap a Dirac-type of topological surface state (TSS) then exists. So far, however, $BaBiO_3$ has not yet experimentally been verified as a topological insulator.

The central question, which we want to address in this work, is to study to what extent the decisive role of the SOC of the $s$- and $p$-Bismuth orbitals for the band inversion and the TI nature can be carried over to the large family of superconducting, i.e. doped, Bismuth Oxides. Here of particular interest are the potassium (K) and lead (Pb)-doped "relatives" of $BaBiO_3$. As the most experimentally studied doped

[1]Institut für Theoretische Physik und Astrophysik, Universität Würzburg, 97074 Würzburg, Germany. [2]Max Planck Institute for Chemical Physics of Solids, 01187 Dresden, Germany. [3]Max Planck Institute for the Physics of Complex Systems, 01187 Dresden, Germany. Correspondence and requests for materials should be addressed to G.L. (e-mail: gangli@physik.uni-wuerzburg.de)

BaBiO$_3$ compounds, these systems are naturally the first choice to understand the influence of doping on the topological nature of BaBiO$_3$. It has been known for a long time, that doping cubic BaBiO$_3$ with K and Pb will convert this system to a superconductor[5,6](SC). Indeed, Ba$_{1-x}$K$_x$BiO$_3$ and BaPb$_{1-x}$Bi$_x$O$_3$ show the highest transition temperatures in copper- and iron-free systems. Thus, the key question here is, whether the doping of BaBiO$_3$ with K and Pb, *i.e.* extending the systems to the hole-doped high-Tc bulk superconductors, will still preserve a "hidden" topological insulator phase achievable through *additional* electron- or also hole-doping. Our work answers this crucial question: for the experimentally known structures and phases of the K- and Pb-doped BaBiO$_3$ compounds, the topological nature is indeed found to be robust, providing the feasibility of doping, which still keeping the topological nontrivial band structure. This should encourage experimentalists to tune BaBiO$_3$ by first achieving an appropriate doping and then performing electric gating on top of this.

These questions are answered via a theoretical study of the other two end compounds, *i.e.* KBiO$_3$ and BaPbO$_3$, in their cubic phase. Additionally, we consider the doped compounds Ba$_{0.5}$K$_{0.5}$BiO$_3$, as well as BaPb$_{0.7}$Bi$_{0.3}$O$_3$, which facilitates an interpolation in our search for 3D TIs within the Ba$_{1-x}$K$_x$BiO$_3$ and BaPb$_{1-x}$Bi$_x$O$_3$ families.

We find that cubic KBiO$_3$ has a very similar electronic structure as BaBiO$_3$, including again the band inversions at the R-point. However, compared to the band inversion in BaBiO$_3$, this happens at a higher energy in KBiO$_3$. Thus, cubic KBiO$_3$ can also become a TI via electron doping, albeit experimentally this may harder to be realized. In BaPbO$_3$, the Bi *s*-band moves to higher energy and while a direct energy gap is left at the R-point, preserving the topological nature[1], the indirect gap is zero. As a result, in the surface BZ the Dirac cone merges into the bulk bands. Under electron doping, BaPbO$_3$ can then be tuned into a "topological metal". As an important additional observation, we find that all three Bismuth Oxides, *i.e.* BaBiO$_3$, KBiO$_3$ and BaPbO$_3$ are found to contain another band inversion at the Γ-point *below* the Fermi level. For this reason, these systems can, in principle, also be tuned into TIs via hole doping. These results concerning the topological nature and the "hidden" Dirac cones are also carried over to arbitrary doping levels, as discussed in our work.

## Results

Let us first briefly review the structural aspects of all three parent compounds and explain why we will focus in this paper on their cubic phase. One striking feature in BaBiO$_3$ is the breathing-mode distortions, induced by Bi ions. It gives rise to an ordered arrangement of Bi$^{3+}$ and Bi$^{5+}$ ions[7]. Comprehensive crystallographic studies, by using neutron powder diffraction[8,9], found that BaBiO$_3$ experiences a number of temperature-induced phase transitions. They range from monoclinic in P2$_1$/n at low temperature and monoclinic in I2/m at room temperature to rhombohedral in R$\bar{3}$ at *ca* 405 K and cubic in Fm$\bar{3}$m at *ca* 750-800 K. In ref.1, the interesting topological phase was found to exist in the perovskite lattice (see crystal structure in Fig. 1(b)) of the parent compound BaBiO$_3$ in the cubic phase, which is stable against the monoclinic lattice distortion. The perovskite structure of BaPbO$_3$ is also quite stable[10,11]. At temperatures above 673 K, it crystallizes in the simple perovskite structure in Pm$\bar{3}$m and transforms to tetragonal I4/mcm structure at temperatures below it. Further decreasing the temperature to 573 K, it changes to the orthorhombic Ibmm structure. KBiO$_3$ does not form the perovskite structure but rather crystallizes in a cubic KSbO$_3$-type tunnel structure with space group Im$\bar{3}$[12].

Despite the large variations in their crystal structures, as far as *superconductivity* is concerned, the Ba$_{1-x}$K$_x$BiO$_3$ compounds are found to confine to their cubic symmetry[13–15]. For this reason, we restrict our study of the parent compound KBiO$_3$ to the simple Pm$\bar{3}$m perovskite structure (see Fig. 1).

The strategy here is as follows: if the two end parent compounds are both TIs in the cubic structure, their random alloys with the same structure would have a large chance to be also TIs. Through first-principle calculations of a superstructure, we confirm this indeed to be the case for Ba$_{1-x}$K$_x$BiO$_3$. Superconducting BaPb$_{1-x}$Bi$_x$O$_3$ is confined to the tetragonal distortion of the cubic phase[16]. Similarly to Ba$_{1-x}$K$_x$BiO$_3$, here we will also concentrate on the cubic phase of BaPbO$_3$ and BaPb$_{1-x}$Bi$_x$O$_3$. In addition, we also show that the tetragonal distortion does not change the topological nature of the cubic BaPbO$_3$.

**KBiO$_3$ and Ba$_{0.5}$K$_{0.5}$BiO$_3$.** Figure 1 displays the electronic structure of simple cubic KBiO$_3$ with the lattice constant 4.2886 Å[13,17]. Its band structure is very similar to that of BaBiO$_3$. Thus, the discussions here also apply to BaBiO$_3$. Cubic KBiO$_3$ is a metal, with a band carrying a large weight of the Bi *s*-orbital crossing the Fermi level. The potassium states stay at higher binding energy and are not relevant to the band-inversions we discuss here. The bands displayed in Fig. 1(a) are mainly from the bismuth and oxygen states. The band-inversions of KBiO$_3$ are shown in Fig. 1(a) as the interchange of the red and green colors at special TRIM (see, for example, the light-yellow area around the R-point). The red and green colors are used to label Bi *s*- and *p*-orbitals, respectively. The interchange of the two colors inside the light-yellow area indicates that the order of these two bands is inverted. Due to this reason, the adiabatic connection to the atomic limit of the system is then lost, which makes this system topologically different from the conventional insulator. An important point for the more general TI physics in these compounds is that, the topological gap shown in the light-yellow area at the R-point is induced by the strong SOC of Bi, as discussed in ref.1 for BaBiO$_3$.

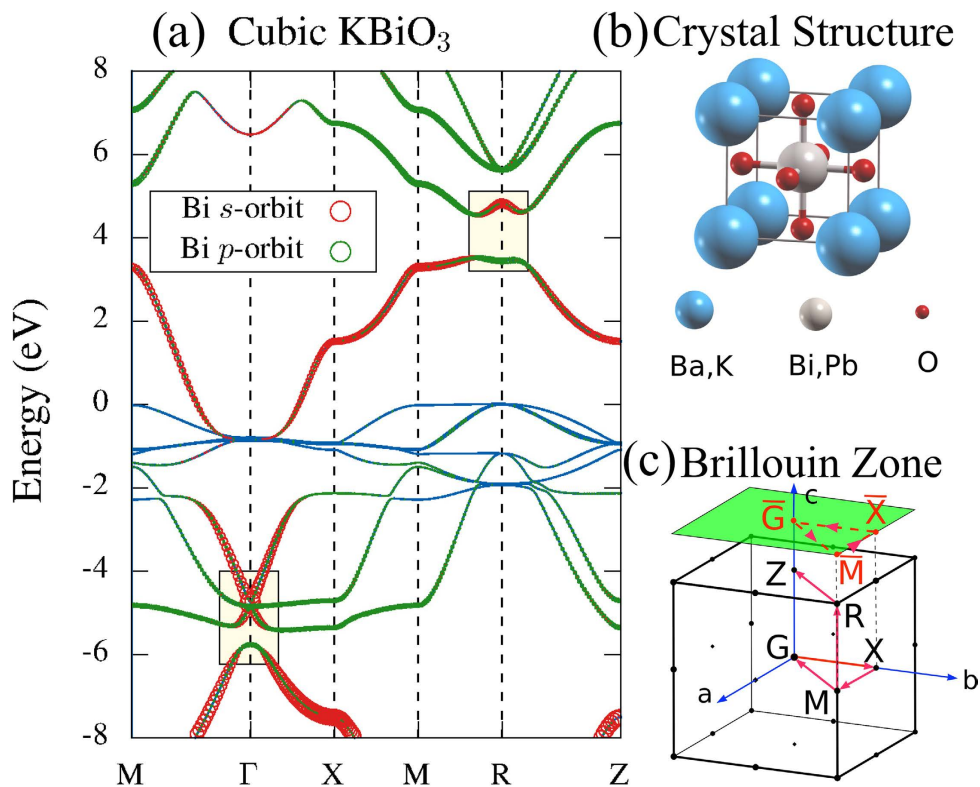

**Figure 1. Electronic structure of KBiO$_3$:** (**a**) along high-symmetry paths in the first BZ, as indicated by the red arrows in (**c**) for the crystal structure (**b**). The width of the red and green colored bands shows the weight of Bi $s$- and $p$-orbitals, the two light-yellow areas mark the band-inversions of this system.

In addition, we found a band inversion at the $\Gamma$-point, however, *below* the Fermi level. Similar to the band inversion at R, the Bi $s$-orbital and $p$-orbital weights interchange at $\Gamma$ around $5.8\,eV$ below $E_F$. It may experimentally not be easy to shift $E_F$ into such heavily hole-doped region. However, it should be feasible for ARPES to detect the corresponding Dirac-type surface states in the occupied bands.

To further validate the topological nature of cubic KBiO$_3$ at both hole- and electron-doped regimes, we calculate the surface states by constructing a slab geometry along the [001] direction of bulk KBiO$_3$. Fig. 2(a, b) displays the electronic structure of such a slab with thickness of 60 atomic layers. As one can clearly see from Fig. 2(a), there are two Dirac cones located at $\overline{\Gamma}$ and $\overline{M}$. They stay at different sides of the Fermi level. Though, the energy gap below $E_F$ is smaller than that at the R-point (above $E_F$), it is clear that there is also a Dirac-cone positioned inside it. Thus, both topological surface states can, in principle, be accessed via "chemical engineering", see the schematic plot in Fig. 3.

By calculating the projected density of states, we, further, verified that both Dirac cones are stemming from surface states (results are not shown here). With the current slab setup displayed in Fig. 2, each branch of the surface states is doubly degenerate as this slab contains two surfaces with the same atomic layer, which do not interact with each other. This degeneracy can be removed by terminating one surface with a different atomic layer, as shown in Fig. 2(b). When one surface is terminated by the Bi-O layer, while keeping the other one unchanged (K-O layer), the different potentials at these two surfaces separate the two Dirac cones in energy, which lifts the degeneracy. Furthermore, the $Z_2$ topological invariant calculated in Table 1 also confirms KBiO$_3$ to be a strong topological insulator.

Since both cubic BaBiO$_3$ and KBiO$_3$ can be TIs, we next show that their *random alloys in the same cubic phase*, in principle, can be also TI. Ba$_{1-x}$K$_x$BiO$_3$ is found to crystallize in the Pm$\overline{3}$m phase for $0.3 < x < 0.5$[13]. By taking the experimental lattice constant $a = 4.2618$ Å[13] and setting $c = 2a$ (relaxation on c does not change the following conclusion), we calculated the electronic structure of Ba$_{0.5}$K$_{0.5}$BiO$_3$ by using a supercell with two Bi atoms (see Fig. 4 for the corresponding crystal structure and the BZ). Comparing with the cubic structure of BaBiO$_3$ and KBiO$_3$, the length of the c-axis is doubled. This leads to a folding of the BZ and to a mapping the R-point of the cubic BZ (see the BZ in Fig. 1) to the M-point. So does the band-inversion.

We found both band-inversions in cubic BaBiO$_3$ and KBiO$_3$ to exist in their corresponding superstructure. Now they are at the TRIM M-point and $\Gamma$-point. Here, we adopted the case $x = 0.5$ as an example to illustrate the surface states. This case allows us to work with a superlattice that is only twice the size of the cubic cell. As displayed in the right plot of Fig. 4, in a slab of Ba$_{0.5}$K$_{0.5}$BiO$_3$ constructed with the superlattice bulk band structure along the [001] direction, these two band-inversions induce Dirac cones located at the $\overline{M}$ and $\overline{\Gamma}$ points at the surface BZ. Thus, substituting Ba with K does not change the topological nature of the system, see also Table 1 for the topological invariant. The band inversion, as well the Dirac cones are robust to this substitution, as long as the alloys crystallize in the cubic structure.

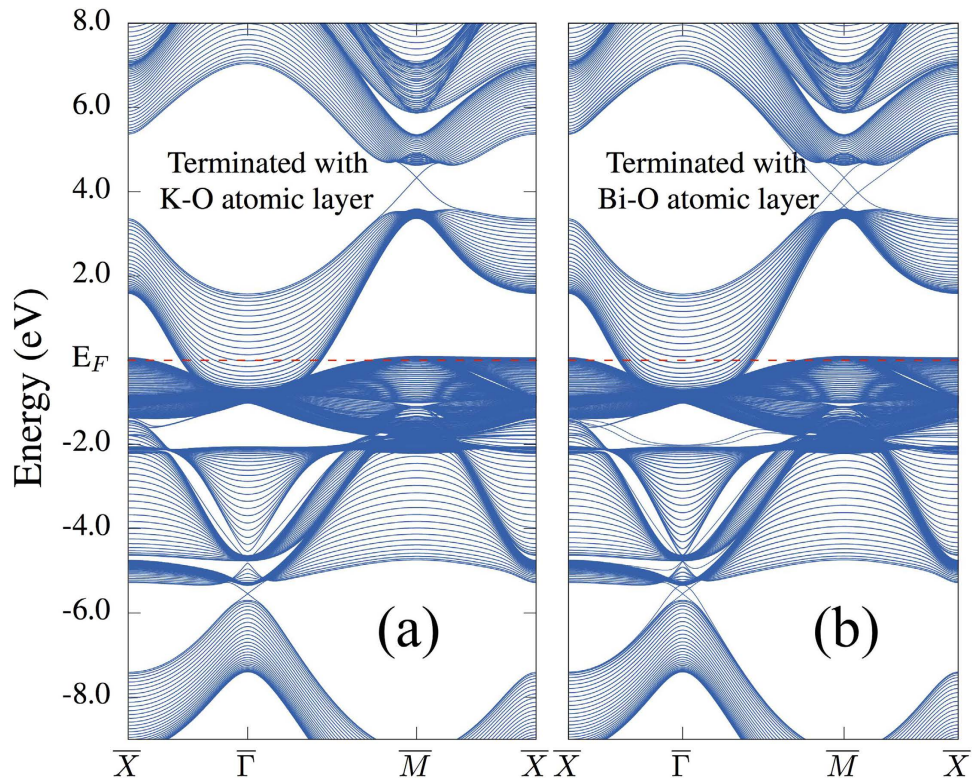

**Figure 2. The surface states of cubic KBiO$_3$ with two different types of surface terminations**: (**a**) the upmost surface is K-O; (**b**) the upmost surface is Bi-O. In both setups, the bottom surface is taken as K-O. Both the top and bottom surfaces can hold Dirac surface states, and the corresponding Dirac cones are located at slightly different energies. The surface BZ is indicated by the green area in Fig. 1(c).

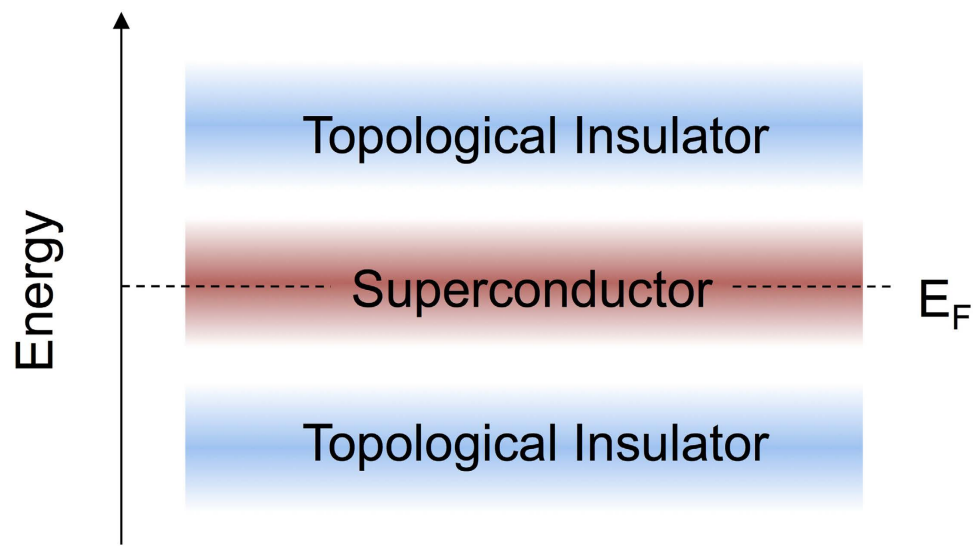

**Figure 3.** A schematic plot indicating the coexistence of the topological insulating and the superconducting phase versus energy in the Bismuth Oxide family.

|  | For the topological gap above $E_F$ | $Z_2$ | For the topological gap below $E_F$ | $Z_2$ |
|---|---|---|---|---|
| KBiO$_3$ (Pm$\bar{3}$m) | 1Γ(+), 3X(+), 3M(+), 1R(−) | (1;111) | 1Γ(−), 3X(−), 3M(−), 1R(+) | (1;111) |
| BaK[BiO$_3$]$_2$ (P4/mmm) | 1Γ(+), 2X(+), 1M(−), 1Z(+), 1A(+), 2T(+) | (1;110) | 1Γ(−), 2X(+), 1M(+), 1Z(−), 1A(−), 2T(+) | (1;110) |
| BaPbO$_3$ (Pm$\bar{3}$m) | 1Γ(+), 3X(+), 3M(+), 1R(−) | (1;111) | 1Γ(−), 3X(−), 3M(−), 1R(+) | (1;111) |
| Ba$_2$PbBiO$_3$ (P4/mmm) | 1Γ(+), 2X(+), 1M(−), 1Z(+), 1A(+), 2T(+) | (1;110) | 1Γ(+), 2X(+), 1M(−), 1Z(−), 1A(−), 2T(−) | (1;110) |
| BaPbO$_3$ (I4/mcm) | 1Γ(−), 2X(+), 4N(+), 1Z(+) | (1;000) | 1Γ(+), 2X(−), 4N(+), 1Z(−) | (1;000) |

**Table 1. Topological invariants for all five compounds studied in this work.** For all states below the topological gaps at above and below the Fermi level, the products of their parity eigenvalues are shown at all eight time-reversal invariant momenta. The $Z_2$ topological invariant then verifies that they are all strong TIs.

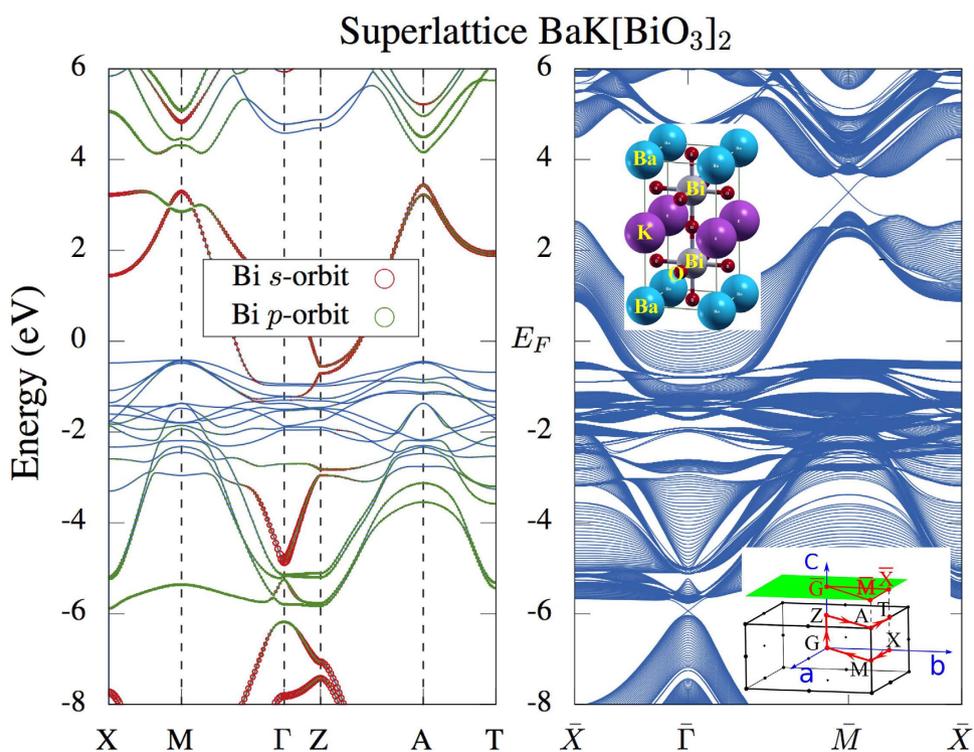

**Figure 4.** The bulk and surface electronic structures of a superstructure composed out of BaBiO$_3$ and KBiO$_3$.

If we, further, compare the location of the energy gap at the R-point in three different systems, *i.e.* BaBiO$_3$, KBiO$_3$ and the superlattice BaK[BiO$_3$]$_2$, the potassium doping effect becomes obvious: K contains less valence electrons compared to Ba. Thus, substituting Ba with K effectively dopes holes into the system. As a result, the electronic structure of KBiO$_3$ can be essentially understood as that of BaBiO$_3$ with a global shift to a larger positive value in energy.

**BaPbO$_3$ and BaPb$_{0.7}$Bi$_{0.3}$O$_3$.** The Bi s-p band inversions in BaBiO$_3$ are left unaffected with the substitution of Ba with K, as the Ba and K states all stay at binding energies outside of the relevant energy regime for this band order reversal. Thus, it is not surprising that their alloys Ba$_{1-x}$K$_x$BiO$_3$, in principle, also stay as TIs in the cubic phase. However, this is not the case in the BaPb$_x$Bi$_{1-x}$O$_3$ systems. Chemically, Pb is adjacent to Bi in the periodic table, with also a large atomic number and, thus, a strong SOC (but weaker than that of Bi). A similar topological phase can then be expected.

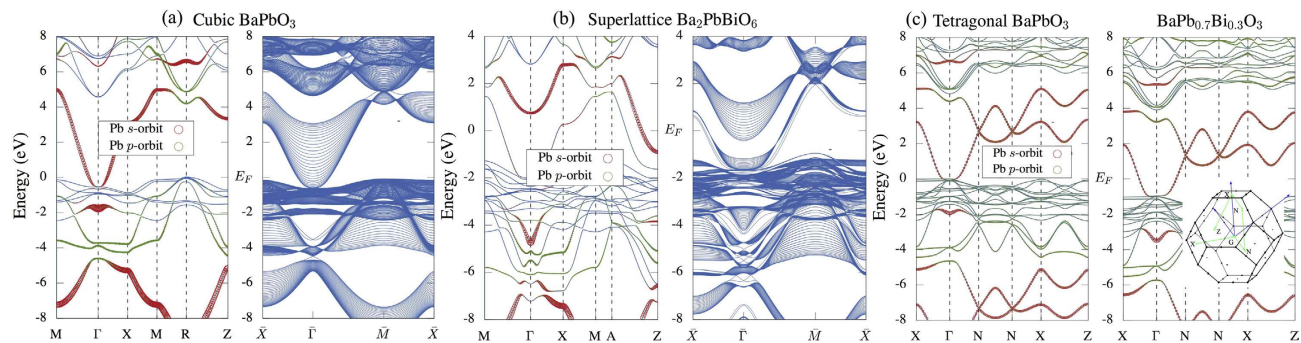

**Figure 5.** The electronic structure of BaPbO$_3$ in the cubic (a) and tetragonal phases (c). The corresponding surface states in a [001] slab are presented in Figs. 5(a, b) in blue color. (**b**) shows the calculations for a superstructure of Ba$_2$PbBiO$_6$, which qualitatively displays the substitution effect of Bi with Pb in the random alloy BaPb$_x$Bi$_{1-x}$O$_3$.

From the following discussions, we will see that the relatively stronger SOC of Bi is crucial for realizing the topological phase in BaBiO$_3$. By substituting Bi with Pb, the reduced SOC leads to a strong shift of the Pb s-type band, which, in the end, pushes the spin-filtered surface states into the bulk states.

At temperatures above 673 K, BaPbO$_3$ crystallizes in the same cubic structure as BaBiO$_3$. Fig. 5(a) displays the electronic structures for the bulk and an [001] slab of cubic BaPbO$_3$. The lattice constant is 4.2997 Å, taken from ref.11. Replacing Bi with Pb strongly modifies the band-inversions at R and Γ points. On the one hand, the band-inversion in cubic BaPbO$_3$ moves to an energy of ~2 $eV$ higher than that in BaBiO$_3$. This is similar to what happens in KBiO$_3$, *i.e.* Pb contains less valence electrons than Bi. Thus, replacing Bi with Pb effectively dopes the system with holes, which shifts the Fermi level downwards.

On the other hand, compared to BaBiO$_3$ and KBiO$_3$ (see Fig. 1), the details of the bulk electronic structure are also strongly modified. The *s*-type band moves to higher binding energy. At R around 5 $eV$, the *p*-type bands shows stronger dispersion along R-M and R-Z directions and the indirect energy gap here becomes negative. Thus, it is of no surprise to see that, when cubic BaPbO$_3$ is terminated with a surface, the surface states originating from this band-inversion merge into the bulk states (see the blue-colored plot of Fig. 5(a)). With the tetragonal distortion (Fig. 5(c)), the band-inversion around 5 $eV$ is still preserved (see also Table 1 for the topological invariant), which now appears at Γ. The separation of the Pb *s*-type and *p*-type bands in energy is also similar to that in the cubic phase (see Fig. 5(a)). For the same reason, the Dirac cone of the tetragonal BaPbO$_3$ and BaPb$_{0.7}$Bi$_{0.3}$O$_3$ also merge into the bulk bands but with their topological nature preserved.

In the right-hand plot of Fig. 5(c), the electronic structure of the tetragonal alloy BaPb$_{0.7}$Bi$_{0.3}$O$_3$ is studied by using the virtual cluster approximation. *This is exactly the phase where superconductivity is observed in experiments.* The band inversion survives the tetragonal distortion and the chemical substitution. Thus, one may be tempted to speculate about topological superconductivity emerging from the coexistence of the superconductivity at T < T$_c$ and the topological insulating nature. It is of crucial importance, however, to note that while the topological insulating phase can indeed be achieved on the basis of our electronic structure calculations, one is actually comparing two different systems (see also Fig. 3): one stands for the doped "high-T$_c$" superconductor with the chemical potential E$_F$ as given by our calculations. The other, *i.e.* the nominal TI system, requires an *additional* shift in the chemical potential to the energies, where the band inversion takes place. However, we find that, in both K- and Pb-doped BaBiO$_3$, the topological gaps below the Fermi level are present. Particularly, in the Pb-doped BaBiO3, the Pb substitution of Bi displays clearly different influences on the topological gaps at above and below the Fermi level. The band-inversion and the energy gap *below* the Fermi level turns out to be more robust, which gives rise to surface states displayed in the right-hand plots in Fig. 5(a) and (b), respectively. If the two surfaces of the slab are terminated with different atomic-layers, one of the Dirac cones will merges into the bulk states, while the other one still is positioned inside the energy gap (see again the surface states plots in Fig. 5 (a) and (b)).

## Conclusions

In summary, it is a fascinating observation to explore topological band structure features "hidden" in the family of high-Tc superconductors, surrounding the BaBiO$_3$ compounds. We found (1) the existence of an additional topological gap and the Dirac cone below the Fermi level, which was not discovered in the previous work[1]. This largely enriches the possibility of observing the Dirac cone in experiment. It should be possible to detect this lower Dirac cone in ARPES; (2) the topological nature of BaBiO$_3$ is robust with respect to K- and Pb-doping. We believe our work represents one crucial step forward towards the final experimental realization of the topological phase of BaBiO$_3$. The robustness of the topological nature of $BaBiO_3$ can make the electric gating much easier in electron doped-$BaBiO_3$ than in the parent compound.

The topological insulator properties and the corresponding Dirac cones embedded within a band gap appear at a different chemical potential than the superconducting state. However, they coexist in the same material class. The SC and the TI states may be related via chemical doping, or other means (see Fig. 3). Our results demonstrate in particular, that, apart from the chemical potential shift, the overall electronic structure remains essentially unchanged under doping, like in a "rigid band structure" description. This observation substantially enhances the possibility of switching from one unconventional state to the other. In that sense, our findings may open up new avenues for the above-mentioned possible applications, such as the fabrication of controlled SC/TI interfaces[31].

## Methods

The calculations were mainly carried out within the full-potential linearized augmented plane-wave (FP-LAPW) method[18], implemented in the package WIEN2k[19]. The package ELK (http://elk.sourceforge.net) and the Vienna Ab Initio Simulation Package (VASP)[20–23] with PAW potentials[24,25] are also employed for comparison. $K_{max}R_{MT} = 9.0$ and a $10 \times 10 \times 10$ k-mesh were used for the ground-state calculations in WIEN2k. $R_{MT}$ represents the smallest muffin-tin radius and $K_{max}$ is the maximum size of reciprocal-lattice vectors. The spin-orbit coupling is included by a second variational procedure. The generalized gradient approximation (GGA) potential[26,27] is used in all calculations. The $Z_2$ invariant for all parent compounds are calculated within WIEN2k.

The surface electronic structures are further calculated using the maximally localized Wannier functions (MLWFs)[28], employing the WIEN2WANNIER[29] and VASP2WANNIER90[30] interfaces. The MLWFs are constructed in a non-self-consistent calculation with an $8 \times 8 \times 8$ k-mesh.

### Acknowledgements

G. Li wants to acknowledge fruitful discussions with J.P., Hu., G. Li and W. Hanke acknowledge the DFG Grant Unit FOR1162 and SPP Ha 1537/24-2. R. Thomale is supported by ERC-TOPOLECTRICS-StG-336012.

### Author Contributions

G.L. designed and led the research, carried out calculations and analysis, documented the findings and prepared the figures; G.L. and W.H. wrote the main manuscript text. B.Y. and R.T. participated in scientific discussions. All authors contributed to the interpretation of results and to the finalization of the submitted manuscript.

### Additional Information

**Competing financial interests:** The authors declare no competing financial interests.

**How to cite this article**: Li, G. *et al.* Topological nature and the multiple Dirac cones hidden in Bismuth high-Tc superconductors. *Sci. Rep.* **5**, 10435; doi: 10.1038/srep10435 (2015).